# Cascaded high-gradient terahertz-driven acceleration of relativistic electron beams

Hanxun Xu[1,2], Lixin Yan[1,2], Yingchao Du[1,2], Wenhui Huang[1,2], Qili Tian[1,2], Renkai Li[1,2], Yifan Liang[1,2], Shaohong Gu[1,2], Jiaru Shi[1,2] and Chuanxiang Tang[1,2]

**Terahertz (THz)-driven acceleration has recently emerged as a new route for delivering ultrashort bright electron beams efficiently, reliably, and in a compact setup. Many THz-driven acceleration related working schemes and key technologies have been successfully demonstrated and are continuously being improved to new limits. However, the achieved acceleration gradient and energy gain remain low, and the potential physics and technical challenges in the high field and high energy regime are still under-explored. Here we report a record energy gain of 170 *keV* in a single-stage configuration, and demonstrate the first cascaded acceleration of a relativistic beam with a 204 *keV* energy gain in a two-stages setup. Whole-bunch acceleration is accomplished with an average accelerating gradient of 85 *MV/m* and a peak THz electric field of 1.1 *GV/m*. This proof-of-principle result is a crucial advance in THz-driven acceleration with a major impact on future electron sources and related scientific discoveries.**

[1]Department of Engineering Physics, Tsinghua University, Beijing CN-100084, China. [2]Key Laboratory of Particle and Radiation Imaging, Tsinghua University, Ministry of Education, Beijing CN-100084, China. Correspondence and requests for materials should be addressed to Wenhui Huang (emai: huangwh@mail.tsinghua.edu.cn)

Acceleration and control of high-quality electron beams using electromagnetic waves ranging from radio-frequency (RF) to optical lasers has been a main thread in the development of modern science and technology. Limited by RF-induced plasma breakdown, conventional accelerators operate with relatively low gradients; hence, high-energy machines are costly and time-consuming to build. A higher operating frequency and shorter pulse length are vital for achieving a higher accelerating gradient, since the RF breakdown threshold scales as $E_s \propto \tau^{-1/4} f^{1/2}$, where $E_s$ is the breakdown threshold of the electric field, $\tau$ is the RF pulse length and $f$ is the operating frequency[1,2]. One promising solution to overcome the RF breakdown is to up-convert the RF into the laser frequency regime. Laser-based acceleration schemes have been demonstrated to realize >$GV/m$ accelerating gradients, including plasma wakefield acceleration[3-7] and dielectric laser acceleration[8-11], which are currently under rapid development. Despite tremendous progress in the past several decades, plasma wakefield acceleration still faces physical and technical challenges to deliver more stable and controllable beams. Due to the submicron optical wavelengths, dielectric laser accelerators suffer difficulties in fabrication and are limited to low bunch charges even in the near-infrared range.

Recently, great interest has been focused on THz-driven acceleration as a new approach sharing certain advantages of RF acceleration and laser-based acceleration. The millimeter-scale THz wavelength simplifies the structure fabrication and supports significant charge per bunch. Moreover, the breakdown threshold can be increased to the multi-$GV/m$ level, offering feasible access to future high-gradient accelerators. The last decades have witnessed extensive milestones in THz-driven acceleration, including THz-driven electron guns[12-13], THz-driven acceleration with low-energy beams[14-16] and relativistic beams[17-18]. Previous works have achieved multi-$MV/m$ accelerating gradients with modest THz sources. Operating THz-driven accelerators with a high gradient is now crucial to exploring the potential physics and technical challenges in the high field. Realizing high-energy THz-driven accelerators based on single-stage acceleration is difficult since the required high-power THz sources are extremely challenging. Multistage THz-driven acceleration is assumed to be inevitable to achieve high-energy accelerators based on modest THz sources. With the experimental demonstration of cascaded THz-driven acceleration of low-energy beams[19], realizing the cascaded THz-driven acceleration in the relativistic regime is of great importance to show the full potential of THz-driven acceleration in high-energy regimes, demonstrating the route to future high-energy THz-driven accelerators.

In this paper, we experimentally demonstrate the first cascade acceleration of a relativistic 34.3 $MeV$ beam and achieve a record energy gain, 170 $keV$ in a single stage and 204 $keV$ in a cascaded configuration, from a THz-driven accelerator. Whole-bunch acceleration of a relativistic beam containing $10^7$ electrons are accomplished in two stages of collinear dielectric loaded waveguide (DLW) structures powered by quasi-single-cycle coherent transition radiation (CTR) THz pulses centered at 0.6 *THz*. We exploited the energy variation over a full cycle of the CTR THz pulse and achieved an average acceleration gradient of 85 *MV/m* with a peak THz electric field of 1.1 *GV/m*. This work is an essential advance in THz-driven acceleration, paving the way for future ultra-compact, high-energy THz-driven accelerators.

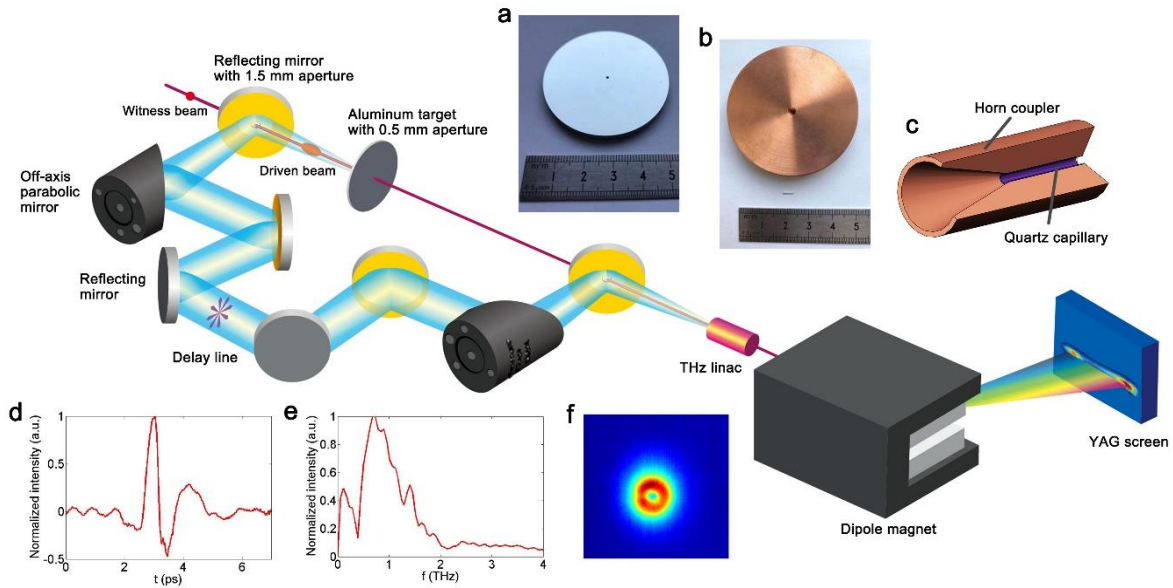

**Figure 1 | Single-stage THz-driven linear accelerator.** An 850 *pC*, 30.4 *MeV* driven beam perpendicularly strikes an aluminum target, thus generating a single cycle CTR pulse centered at 0.6 THz. The radially polarized THz pulse couples into the DLW via a tapered horn coupler and interacts with the following 1.9 *pC*, 34.3 *MeV* witness beam. Timing between the electron and THz wave can be configured by the finely adjusting optical delay line. The energy spectrum of the witness beam is determined via a dipole spectrometer located at the end of the beam line. Inset: **(a)** Aluminum target. **(b)** Horn coupler and quartz capillary. **(c)** Schematic of the THz linac. **(d)** Temporal and **(e)** spectral profiles of the CTR THz pulse determined by electro-optical sampling measurements. **(f)** Focal spot profile of the CTR pulse.

## Results

**Single-stage THz acceleration.** A schematic of the single-stage THz-driven linear accelerator is shown in Fig. 1. The Tsinghua Thomson scattering X-ray source (TTX)[20-21] beamline is configured to provide two electron beams (see Methods: Double bunch generation), an 850 *pC*, 30.4 *MeV* driven beam for generating the CTR pulse and a 1.9 *pC*, 34.3 *MeV* witness beam following the driven beam for the THz-electron interaction. The driven beam perpendicularly strikes an aluminum target, generating a 132 *µJ* CTR THz pulse centered at 0.6 *THz* (see Methods: Double THz pulse generation and characterization), as depicted in Fig. 1d and Fig. 1e. The natural radially polarized THz pulse (see Fig. 1f), is coupled to a DLW via a tapered horn coupler, thus exciting a traveling TM01 mode electromagnetic field inside the waveguide. The longitudinal electric field component of the TM01 mode accelerates the witness beam when the proper timing is fixed. The bunch length of the witness beam is estimated to be 460 *fs* FWHM (see Methods: Double bunch generation), which is shorter than the half-cycle of the THz field; hence, a whole-bunch acceleration can be expected. The THz linac is designed to maximize the energy gain, optimize the field distribution and improve the wave coupling (see Methods: THz linac). Fig. 1b and Fig. 1c show a schematic of the THz linac. The inner diameter of the DLW is 436 *µm* with a dielectric wall thickness of 52 *µm*. The linac is 10 *mm* in length, including a 5 *mm* tapered horn for THz wave coupling. The inner diameter of the CTR target is 0.5 *mm*, which is sufficiently large for the witness beam to transmit while the target intercepted most of the driven beam, generating a significant CTR pulse (see Fig. 1a). Two reflecting mirrors with a 1.5 *mm* hole in the center are employed to reflect the CTR pules while allowing the witness beam to transmit. The witness beam penetrates through a series of small apertures and finally enters the waveguide and copropagates with the THz wave. The THz-electron timing can be configured by finely adjusting the optical delay line. Nevertheless, the timing jitter is expected to be negligible since the THz pulse and the witness beam are both produced by the same driven laser. The energy spectrum of the witness beam is determined via a dipole spectrometer located at the end of the beam line. Optically-generated THz sources are also suitable for driving our THz linac after converting the linearly polarized THz wave into a radially polar-ized wave (see Methods: Double THz pulse generation and characterization).

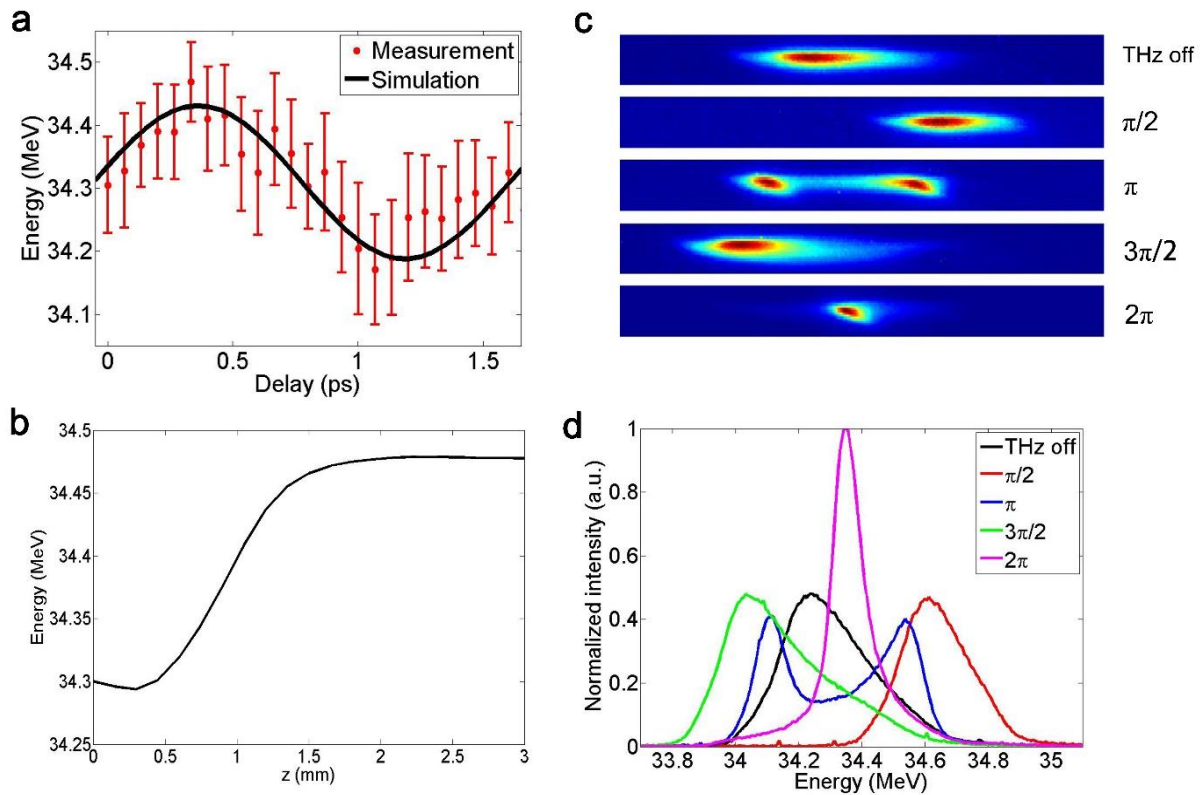

**Figure 2 | Single-stage THz acceleration and energy modulation. (a)** Mean energy of the witness beam as a function of the THz pulse delay. Red dots are measured values with one s.d. error bars over 100 shots, and the black line is the simulation result. **(b)** Simulation of the experimental setup when the injection phase is set to π/2. **(c)** Snapshots of the energy spectrometer at different injection phases. **(d)** Normalized projected energy spectra corresponding to (c).

Fig. 2a shows the mean energy of the witness beam as a function of the THz pulse delay. The curve is compared with the CST[22] particle-in-cell (PIC) simulation result used to model the THz linac. The energy varies over the full $2\pi$ range of the injection phase with a maximum energy gain of 170 $keV$. The DLW is designed to match the phase velocity of the witness beam at 0.6 $THz$. Thus, the phase velocity of the THz wave equals the speed of the witness beam (0.9999$c$), while the group velocity is 0.48$c$. A temporal walk-off arises when the electron beam loses overlap with the THz wave envelope since the electron velocity is higher than the THz group velocity, resulting in a limited THz-electron interaction length. At the maximum accelerating phase, the simulation shows a 2 $mm$ THz-electron interaction length (see Fig. 2b), and the effective accelerating gradient of the THz linac is calculated to be 85 $MV/m$. The corresponding peak electric field in the DLW is estimated to reach 1.1 $GV/m$. Four key phase points are investigated in Fig. 2c and Fig. 2d and compared with the initial energy spectrum. The single energy peak shifts to the right (left) when the injection phase is set to $\pi/2$ ($3\pi/2$), which contributes unambiguous evidence of the whole-bunch acceleration (deceleration). The witness beam is generated with a negative energy chirp; i.e., the electron energy at the bunch head is higher than that at the bunch tail (see Methods: Double bunch generation). At the $\pi$ injection phase, the energy spectrum splits into two peaks, as the bunch head experiences acceleration while the bunch tail experiences deceleration. The measured energy spread increases from 167 $keV$ initially to 210 $keV$. In contrast, at the $2\pi$ injection phase, an energy dechirp arises when the bunch head is decelerated while the bunch tail is accelerated, resulting in an energy spectrum compression. The measured energy spread is compressed to 141 $keV$, corresponding to a 26 $keV$ energy spread decrease.

**Two-stages THz acceleration.** Fig. 3 illustrates the two-stages THz-driven acceleration experimental setup. A newly designed CTR target with a 50 $nm$ thick tantalum film on a 3 $mm$ thick TPX substrate is applied; This target delivers significant CTR pulses simultaneously on both sides, as the TPX has high transmittance in the 0.2$THz$ ~ 1$THz$ frequency range. Both THz pulses are radially polarized and centered at approximately 0.6 $THz$, as they are generated by the same driven beam striking the same target[23] (see Fig. 3b and Fig. 3c). The THz pulse energy is 66 $\mu J$ (backward) and 35 $\mu J$ (forward). Two THz linacs are separately powered by these backward and forward CTR THz waves. The second stage is a copy of the first stage (the same as described in Fig. 1). By proper phasing of each stage, the witness beam can be accelerated twice by these two THz linacs.

To demonstrate the cascade acceleration of the two-stages THz linacs, we first operate each single-stage THz linac separately. Fig. 4a shows the mean energy of the witness beam as a function of the THz pulse delay compared with the simulation results when linac 1 and linac 2 work individually. A maximum energy gain of 128 $keV$ is achieved in linac 1, while a smaller maximum energy gain of 89 $keV$ is obtained in linac 2 due to the lower pulse energy of the forward CTR pulse. These results reveal that the single-stage linacs can individually work effectively. The measurements shown in Fig. 4c and Fig. 4d verify the cascade acceleration of the witness beam; the energy peak shifts to the right twice when linac 1 and linac 2 begin to operate in sequence. A total energy gain of 204 $keV$ was obtained with a 115 *keV* energy gain in the first stage and an 89 *keV* energy gain in the second stage, which agrees well with the simulation result shown in Fig.4b. The measured bunch charge of the witness beam decreases from 1.9 *pC* initially to 350 *fC* after the second stage. The main reason for particles loss is the collimation tolerance and beam mismatching, which could be mitigated by precisely collimating and employing additional beam focusing components. The total CTR THz pulse energy applied in the cascaded experiment is smaller than that employed in the single-stage experiment (cascaded: 101 *μJ,* single-stage: 132 *μJ*). Nevertheless, the obtained total energy gain is higher than that obtained in the single-stage experiment (cascaded: 204 *keV,* single-stage: 170 *keV*). This comparison reveals the advantage of the cascaded acceleration scheme, which can lower the demand on the required THz energy.

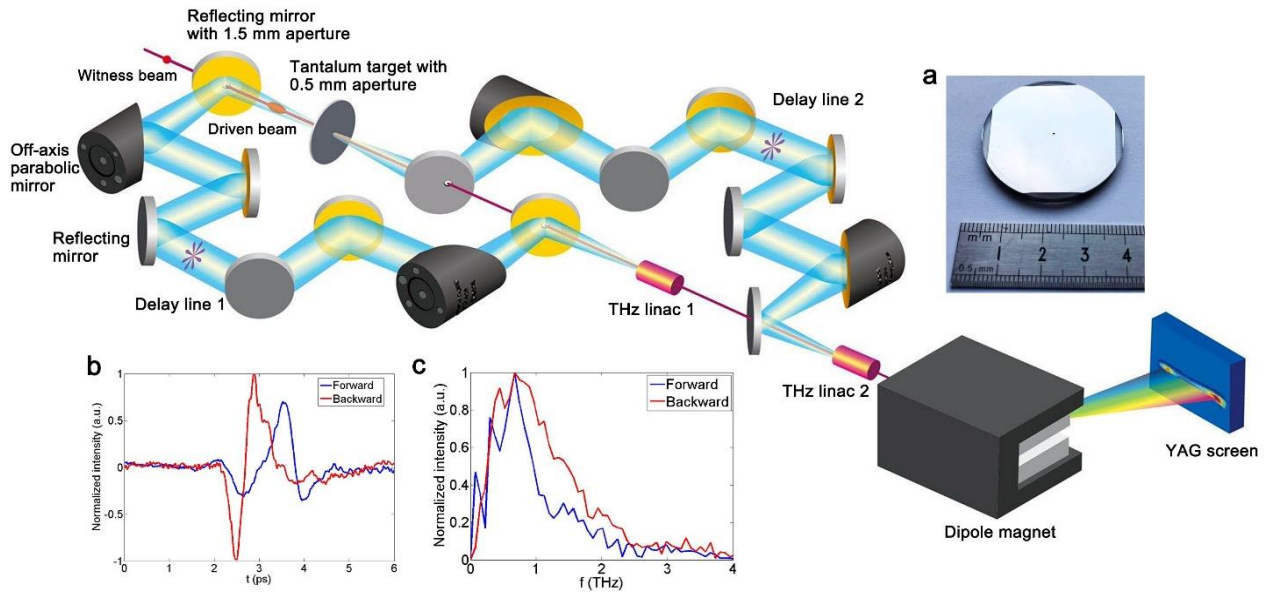

**Figure 3 | Two-stages THz-driven acceleration.** Two THz linacs are separately powered by the backward and forward CTR THz pulses. The second stage is a copy of the first stage (the same as described in Fig. 1). The main difference is the newly designed CTR target enabling significant THz radiation in both the backward and forward directions. Inset: **(a)** The tantalum target. **(b)** Temporal and **(c)** spectral profiles of the CTR THz pulse in backward and forward directions determined by electro-optical sampling measurements.

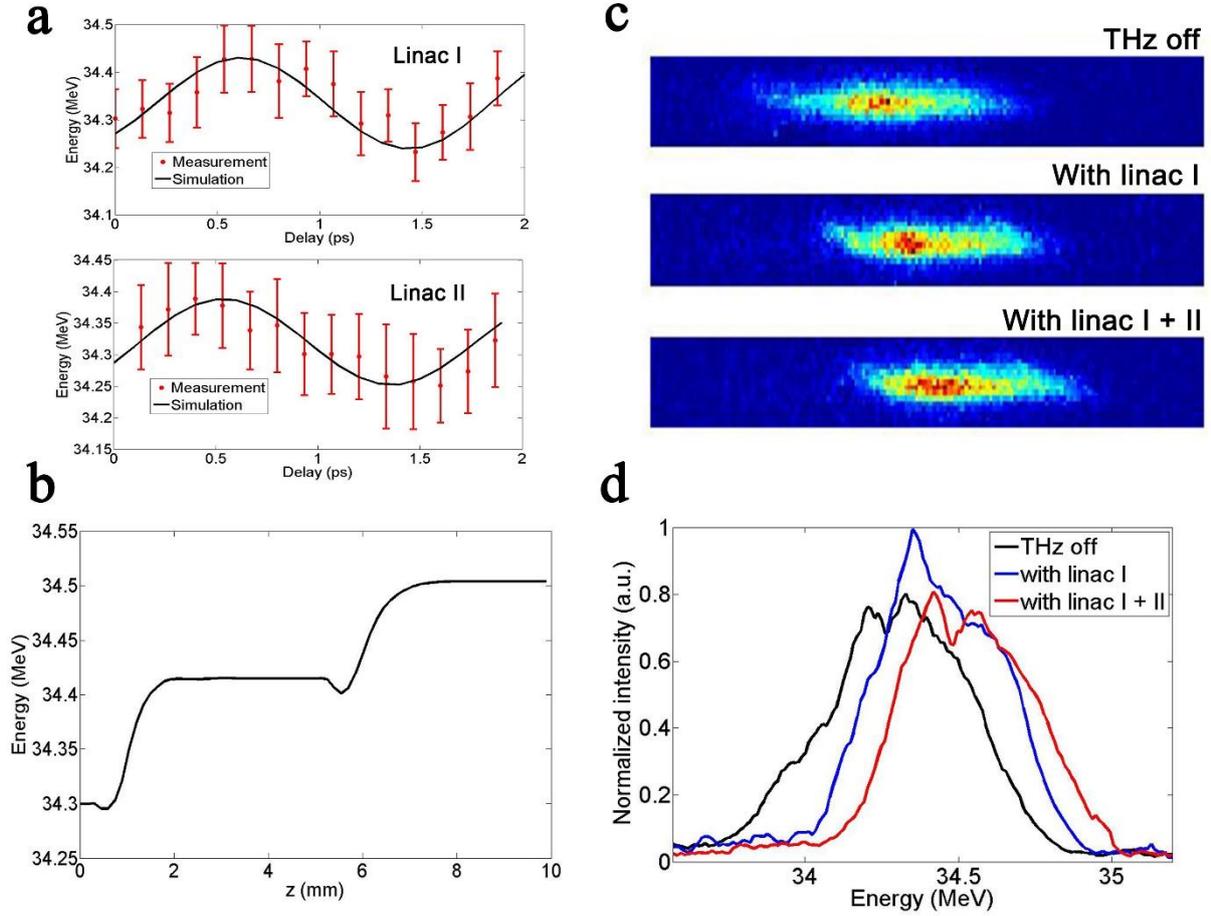

**Figure 4 | Observation of cascaded acceleration. (a)** Mean energy of the witness beam as a function of the THz pulse delay when linac 1 and linac 2 work individually. Red dots are measured values with one s.d. error bars over 100 shots, and the black line is the simulation result. **(b)** Simulation of the cascade acceleration. **(c)** Snapshots of the energy spectrometer when linac 1 and linac 2 begin to operate in sequence. **(d)** Normalized projected energy spectra corresponding to (c).

In summary, we have successfully demonstrated the first cascaded acceleration of a *pC* charge relativistic electron beam with a record energy gain, 170 *keV* in a single stage and 204 *keV* in a cascaded configuration. With the finely optimized THz linac and the hundred-*μJ* level CTR THz pulse, an effective accelerating gradient up to 85 *MV/m* was obtained. Taking advantage of the short bunch length of the witness beam, we achieved whole-bunch acceleration and exploited the energy variation over a full cycle of the THz pulse. Beam energy spectra manipulation is also investigated, resulting in obvious energy spectra stretching and compression.

Currently, high-power THz generation is under rapid development[24-28] and several *mJ* level THz sources are foreseeable. With upgrades to the THz source, dynamic simulation predicts a 1.5 *MeV* energy gain assuming a 10 *mJ* THz source and an initial electron energy of 1 *MeV*. The energy gain can be doubled to 3.1 *MeV* by employing a four-stages acceleration scheme (see Methods: THz linac). In the future, multicycle high-power THz sources can further improve the THz-electron interaction length and the accelerating gradient, and multiple stages of THz-driven acceleration can be applied to achieve higher energy gain with additional THz sources for subsequent stages. With technological improvement to THz acceleration, THz-driven accelerators will play an important role in the development of next-generation compact electron sources.

This proof-of-principle experiment demonstrates the cascaded THz-driven acceleration scheme in the unexplored relativistic energy regime, exploring the beam dynamics in the high field and high-energy regime. This work takes a crucial step in the development of the THz-driven acceleration, showing a feasible route to future ultracompact and high-energy THz-driven accelerators, which will provide users with laboratory-scale electron sources that enable research in ultrafast electron diffraction, X-ray sources and related science.

## Methods

**Double bunch generation.** The driven beam and the witness beam used in this experiment were generated in the TTX beam line at Tsinghua University, as shown in Fig. 5a. Two laser pulses with a preset time delay are produced via a beam splitting technique (see in Fig. 5b). A 9 *ps* flat-top laser pulse was stacked from a 700 *fs* (FWHM) pulse using four *α*-BBO crystals. The other 700 *fs* (FWHM) split pulse was transmitted through an optical delay line with a 1.41 *ns* time delay (approximately 4 RF cycles). The 850 *pC* driven beam and the 1.9 *pC* witness beam were produced by directing the double-pulse ultraviolet (UV) laser to the photocathode in the S-band RF gun. The driven beam was accelerated at the -45° RF phase, resulting in a positive energy chirp for subsequent bunch compression in the magnetic chicane. The measured energies of the driven beam and the witness beam were 30.4 *MeV* and 34.3 *MeV*, respectively, as determined via a dipole spectrometer located at the end of the beam line. A dynamic simulation based on ASTRA[29] showed that the bunch length was compressed to 700 *fs* FWHM (driven beam) and 460 *fs* FWHM (witness beam) after the magnetic chicane. The center frequency of the CTR wave was 0.6 *THz*, corresponding to a 1.67 *ps* wavelength. The bunch length of the witness beam is approximately 1/4 cycle of the CTR pulse, enabling a whole-bunch acceleration or deceleration. The magnetic chicane was configured to minimize the bunch length of the driven beam, thus maximizing the CTR pulse energy. The witness beam was over compressed while the driven beam was optimally compressed, resulting in a negative energy chirp; i.e., the electron energy at the bunch head was higher than that at the bunch tail.

**Double THz pulse generation and characterization.** Transition radiation is emitted when a relativistic beam crosses the boundary between a vacuum and metal. If the radiation wavelength exceeds the electron bunch length, all of the electrons emit coherently. For a 50 *nm* thick tantalum target, an 850 *pC*, 30.4 *MeV* driven beam would generate a considerable CTR wave in both the backward and the forward directions. The pulse energy was measured by a Gollay cell detector, and the focal spot profile was obtained by a THz camera. A single-shot electro-optical sampling method[30] was applied to determine the temporal profile of the CTR pulse. Powering the THz linac with this beam-based CTR THz source provided the following advantages: (a) The initial CTR radiation was radial polarized and coupled well to the TM01 mode of the DLW in the far field. (b) This beam-based THz source benefited natural synchronization since the THz pulses and the witness beam were produced by the same driven laser. (c) CTR radiation can generate a single cycle THz pulse with considerable energy. (d) The center frequency of CTR was tunable in the frequency range of 0.2 *THz* ~ 1 *THz* by varying the bunch length of the driven beam, which can be easily achieved by changing the working current of the magnetic chicane.

Optically-generated THz sources are also suitable for driving our THz linac. High-power, frequency-tunable optically-generated THz sources are under rapid development[24-27]. The polarization of an optically-generated THz wave depends on the driven laser, which is usually linearly polarized. A linearly polarized THz wave can be converted to a radially polarized wave using a segmented waveplate. After converting the polarization, an optically-generated THz source can play the same role as the CTR THz wave in this work.

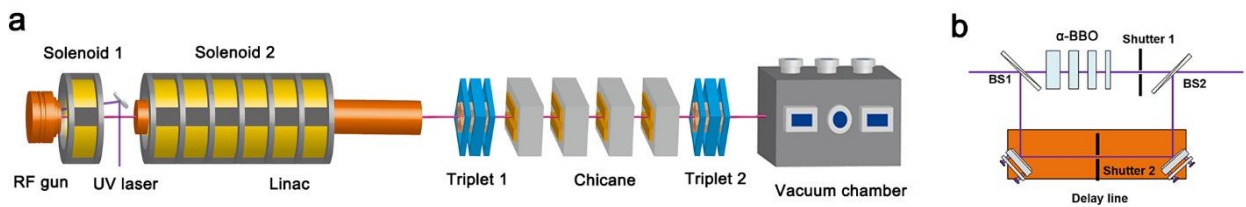

**Figure 5 | Double electron bunch generation. (a)** Schematic of the TTX beam line. The TTX beam line consists of an S-band rf gun, a 3*m* SLAC traveling wave accelerating tube, a magnetic chicane, two solenoids for emittance compensation and two triplets for beam focusing. The THz-driven acceleration setup is located inside the vacuum chamber. **(b)** Schematic of the laser splitting setup. The driven pulse is stacked from a 700 *fs* (FWHM) pulse using four *α*-BBO crystals. The following witness pulse is delayed by an optical delay line. BS: beam splitter.

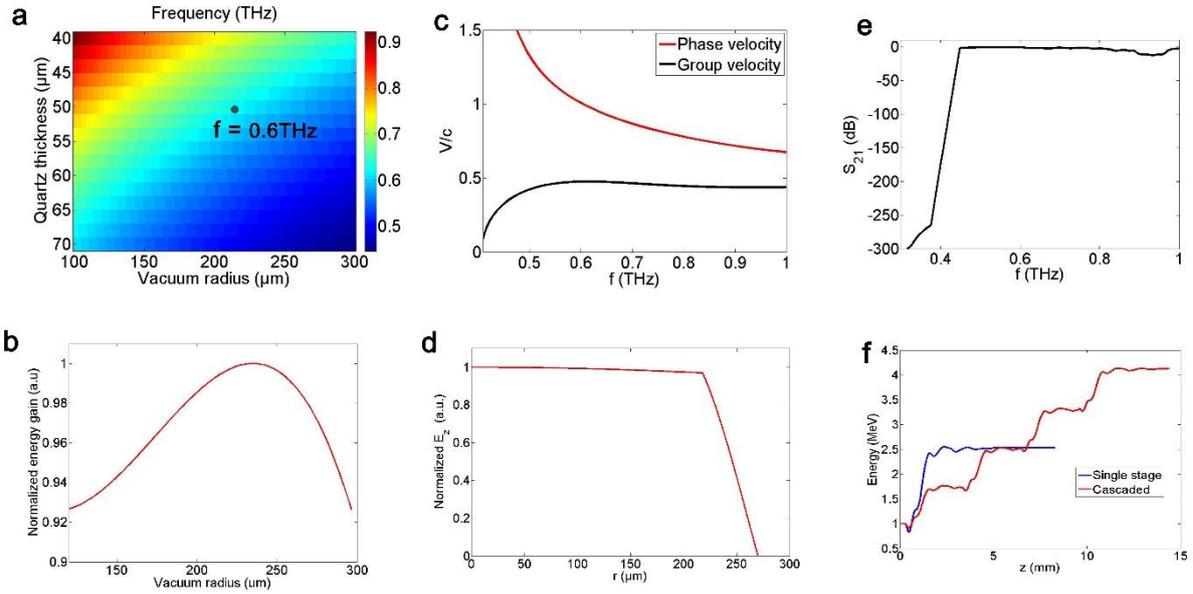

**Figure 6 | THz linac. (a)** Frequency map for different dimensions when $v_p = 0.9999c$. **(b)** Calculated normalized energy gain for different dimensions when $v_p = 0.9999c$ and $f_0 = 0.6$ *THz*. **(c)** Dispersion curve of the designed DLW. **(d)** Normalized magnitude of the longitudinal electric field component along the radial axis. **(e)** Calculated coupling of the horn coupler. **(f)** The electron energy as a function of distance for two case: single stage and cascaded configurations, assuming a 10 *mJ* THz source and an initial electron energy of 1 *MeV*.

**THz linac.** The THz linac consists of a quartz capillary inserted into a hollow cylindrical copper tube with a tapered horn coupler. The permittivity of quartz is approximately 3.85. The DLW is design to match the phase velocity of the witness beam at 0.6 *THz*. Limited by the dispersion relation, for a given vacuum radius and dielectric thickness, there is only one working frequency at which the phase velocity equals the electron speed (0.9999c). Fig. 6a shows the calculated frequency map for different dimensions when $v_p = 0.9999c$. A larger vacuum radius is desirable for a larger group velocity, which will contribute to a longer THz-electron interaction length. However, the peak field along the axis decreases as the vacuum radius increases, which results in a lower accelerating gradient. There is a compromise between increasing the interaction length and strengthening the accelerating field. Fig. 6b shows the calculated energy gain for different dimensions when $v_p = 0.9999c$ and $f_0 = 0.6$ *THz*. The inner diameter of the quartz capillary is 436 *μm* with a dielectric wall thickness of 52 *μm*. The corresponding group velocity is 0.48c, as shown in the dispersion curve in Fig. 6c. The longitudinal electric field is optimized to be fairly homogenous in the vacuum regime with only 3% reduction from the vacuum axis to the dielectric wall (see Fig. 6d). Fig. 6e shows the simulated coupling for the THz wave to the DLW with well coupling over a bandwidth more than 300 *GHz* around the center frequency at 0.6 *THz*. With increased THz energy, Fig. 6f shows the electron energy variation for two cases: single stage and cascaded configurations, assuming a 10 *mJ* THz source and an initial electron energy of 1 *MeV*. Dynamics simulations based on a single particle model predict a 1.5 *MeV* energy gain using a single-stage configuration powered by a 10 *mJ* single-cycle THz pulse. Furthermore, an energy gain of 3.1 *MeV*, nearly doubling the energy gain, is achieved by applying a four-stages acceleration setup, each subsequent stage is powered by a 2.5 *mJ* single-cycle THz pulse. The energy can be further increased by employing a multiple stages acceleration scheme driven by a multicycle high-power THz sources.

## Acknowledgements

We acknowledge professor Wanjun Jiang and Shuaihua Ji for the help with the tantalum target coating. This work was supported by the National Natural Science Foundation of China (NSFC Grants No. 11835004) and Science Challenge Project (No. TZ2018005).

## Author contributions

Hanxun Xu, Wenhui Huang, Lixin Yan, and Yingchao Du conceived and designed the experiment. Hanxun Xu built and conducted the experiment with the help from Qili Tian, Lixin Yan, Yingchao Du, Shaohong Gu, Yifan Liang, Wenhui Huang and Chuanxiang Tang. Hanxun Xu designed the dielectric loaded waveguide and the tapered horn coupler with the help of Jiaru Shi. Hanxun Xu developed and performed the simulations for data evaluation and interpretation of results. Hanxun Xu wrote the manuscript with contribution from Wenhui Huang, Lixin Yan, Renkai Li, Yingchao Du and Chuanxiang Tang. Wenhui Huang provided management and oversight to the project.